\documentclass{emulateapj}

\usepackage{graphicx}
\usepackage{epsf}
\newcommand{\nM}{\frac{dn}{dM}}
\newcommand{\Mobs}{M_{\rm obs}}
\newcommand{\nMobs}{\frac{dn}{d\rm\Mobs}}

\newcommand{\Var}{\mbox{Var}}
\newcommand{\tc}{c'}

\newcommand{\Mbias}{M_{\rm bias}}
\newcommand{\sigmalnM}{\sigma_{\rm \ln M}}
\newcommand{\avg}[1]{\langle#1\rangle}
\newcommand{\mbar}{{\bar m}}
\newcommand{\C}{\bold C}
\newcommand{\D}{\bold D}

\newcommand{\hiMsun}{h^{-1}{M_\odot}}

\newcommand{\OmegaDE}{\Omega_{\rm DE}}

\begin{document}

\journalinfo{The Astrophysical Journal, 688:729--741, 2008 December 1 }
\submitted{Received 2008 March 11; accepted  2008 July 13}
\shortauthors{WU, ROZO, \& WECHSLER}
\shorttitle{EFFECTS OF HALO ASSEMBLY BIAS ON SELF-CALIBRATION}
\title{The Effects of Halo Assembly Bias on Self-Calibration in Galaxy Cluster Surveys}

\author{Hao-Yi Wu \altaffilmark{1}, Eduardo Rozo \altaffilmark{2},
Risa H. Wechsler \altaffilmark{1}}

\altaffiltext{1}{Kavli Institute for Particle Astrophysics and
Cosmology, Physics Department, Stanford Linear Accelerator Center,
Stanford University, Stanford, CA 94305; hywu@stanford.edu,
rwechsler@stanford.edu}

\altaffiltext{2}{The Center for Cosmology and Astro-Particle Physics,
The Ohio State University, Columbus, OH 43210;
erozo@mps.ohio-state.edu}

\begin{abstract}

Self-calibration techniques for analyzing galaxy cluster counts
utilize the abundance and the clustering amplitude of dark matter
halos.  These properties simultaneously constrain cosmological
parameters and the cluster observable--mass relation.  It was recently
discovered that the clustering amplitude of halos depends not only on
the halo mass, but also on various secondary variables, such as the
halo formation time and the concentration; these dependences are
collectively termed ``assembly bias.''  Applying modified Fisher
matrix formalism, we explore whether these secondary variables have a
significant impact on the study of dark energy properties using the
self-calibration technique in current (SDSS) and the near future (DES,
SPT, and LSST) cluster surveys. The impact of the secondary dependence
is determined by (1) the scatter in the observable--mass relation and
(2) the correlation between observable and secondary variables.  We
find that for optical surveys, the secondary dependence does not
significantly influence an SDSS-like survey; however, it may affect a
DES-like survey (given the high scatter currently expected from
optical clusters) and an LSST-like survey (even for low scatter values
and low correlations).  For an SZ survey such as SPT, the impact of
secondary dependence is insignificant if the scatter is 20\% or lower
but can be enhanced by the potential high scatter values introduced by
a highly -correlated background.  Accurate modeling of the assembly
bias is necessary for cluster self-calibration in the era of precision
cosmology.

\end{abstract}


\section{Introduction}

The observed accelerating expansion of the Universe, which is often
interpreted as evidence of dark energy, is one of the most surprising
results of modern cosmology.  In the $\Lambda$CDM paradigm, dark
energy governs the late time expansion of the Universe, halting the
growth of structures.  Consequently, the evolution of the number of
massive galaxy clusters provides one of the most powerful probes of
dark energy
\citep[e.g.][]{Wang98,Haiman01,Holder01,Levine02,Hu03,Majumdar03,Rozo07a,Gladders07,Mantz07}.

Several planned and ongoing surveys will identify massive clusters
over substantial volumes using a variety of techniques, including
optical galaxy counts \cite[e.g.][]{SDSS, DES, LSST}, the
Sunyaev-Zel'dovich effect \cite[e.g.][]{SPT, ACT}, and X-ray emissions
\cite[e.g.][]{Ebeling07MACS,400d}.  These cluster surveys will
complement a variety of future dark energy measurements using tools
such as Type Ia supernovae, weak lensing, and baryon acoustic
oscillations.  Since each of these methods is subject to different
systematics, combining them thus provides cross checks necessary to
avoid biased inferences on the properties of dark energy
\citep{Albrecht06}.

While the abundance of clusters as a function of mass is well
understood from a theoretical standpoint, measuring this abundance
relies on observable mass tracers.  This reliance is the single most
significant obstacle confronting the use of clusters as cosmological
probes.  In particular, the statistical observable--mass relation
needs to be understood to high accuracy in order to avoid systematic
errors in the inference of cosmological parameters. Alternatively,
additional observable quantities that depend on halo mass allow one to
simultaneously constrain cosmology and the aforementioned
observable--mass relation.  One such observable quantity is the
clustering amplitude of clusters, which depends sensitively on mass
and can be determined through a counts-in-cells analysis.  This
general method is often referred to as ``self-calibration''
\citep{Majumdar04,LimaHu04,LimaHu05,LimaHu07}.

In this work, we explore a possible systematic that arises in the
self-calibration analysis, namely, the dependence of the clustering
amplitude of halos on secondary variables.  The clustering amplitude
of halos is characterized by the halo bias, and recent studies have
shown that halo bias depends not only on halo mass but also on
additional halo properties, such as concentration, formation time,
spin, substructure fraction, etc.  \citep[e.g.][]{Gao05, Harker06,
Wechsler06, GaoWhite07, Wetzel07, Bett07, Jing07}.  These dependences
are often interpreted as arising from the different assembly histories
of halos of the same mass, and we refer to these dependences
collectively as ``assembly bias'' \citep[e.g.][]{Croton07}.  If
cluster selection is biased with respect to any of these variables,
the observed clustering amplitude of clusters will deviate from the
mean clustering amplitude of clusters with the same mass distribution.
This deviation will lead to a biased inference of the observable--mass
relation, and therefore to biased estimates for the cosmological
parameters of interest.

We herein take the secondary parameter to be the halo concentration,
which has been shown to play a role in halo bias for massive clusters
by Wechsler et al.\@ (2006; see also \citealt{Wetzel07};
\citealt{Jing07}).  We then incorporate the concentration dependence
of halo bias into the standard self-calibration formalism developed in
\cite{LimaHu04, LimaHu05}.  With modified Fisher matrix formalism, we
investigate the impact of this additional dependence on cosmological
parameter estimates from self-calibration.  We specifically calculate
the expected effects for four example galaxy cluster surveys, which
represent the Sloan Digital Sky Survey (SDSS; assuming clusters
selected from the photometric data), the Dark Energy Survey (DES), the
South Pole Telescope (SPT), and the Large Synoptic Survey Telescope
(LSST).  We also explore various assumptions about the correlation
between cluster observable and concentration.  In detail, the
significance of this systematic effect depends on the strength of this
correlation as well as on the observable--mass scatter.  We find that
the resulting bias in the inferred cosmological parameters is
insignificant for the current SDSS photometric surveys, but it can be
significant for upcoming photometric surveys such as DES and LSST.  On
the other hand, for SZ this systematic is less likely to be
significant if the scatter is small and mainly intrinsic, but may
still be significant if the correlated background dominates the
scatter.

This paper is organized as follows.  In \S\ref{sec:bias} we discuss
why assembly bias may lead to biased cosmological parameter estimates
in cluster counting experiments. In \S\ref{sec:self-calibration} we
review the standard self-calibration formalism, and then proceed in
\S\ref{sec:assemblybias} to include assembly bias into this formalism.
Our statistical methodology for estimating the systematic errors due
to assembly bias is described in \S\ref{sec:fisher}.  Details of our
implementation can be found in
\S\ref{sec:implementation}.  Section \ref{sec:results} presents our results
and discussion.  We summarize in \S\ref{sec:conclusions}.


\section{Halo Bias and Dark Energy: Why Assembly Bias Matters}\label{sec:bias}

Halo bias characterizes the clustering amplitude of dark matter halos,
and it is typically defined as the ratio between the density contrast
of halos and that of the dark matter.  In the hierarchical structure
formation predicted by CDM, halo bias is a strong function of mass,
increasing for more massive halos.  This dependence on mass is now
well calibrated from numerical simulations and can be approximated
analytically with the excursion-set theory
\citep[e.g.][]{MoWhite96,Sheth01, SeljakWarren04, Zentner07}.  Halo
bias depends sensitively on dark energy in a way that is complementary
to the dependence of the mass function on dark energy; thus, including
the halo bias information in cluster counting experiments improves the
dark energy constraints from using mean halo abundances alone.

Much work on halo bias has made the simplifying assumption that halo
bias depends only on halo mass. However, recent studies based on
\emph{N}-body simulations have found evidence that secondary variables such
as halo formation time and concentration do impact halo bias
\citep[e.g.][]{Gao05,Harker06,Wechsler06, GaoWhite07,
Wetzel07,Bett07,Jing07}.  In this work, we focus on the impact of halo
concentration on halo bias, principally because among all secondary
parameters, this dependence is the strongest at cluster scales and is
the best understood statistically.  The halo concentration describes
the halo density profile and is defined as $c = R_{\rm vir}/r_s$, where
$r_s$ is the radius where the density profile has a log slope of $-2$.
The halo concentration has been shown to correlate tightly with the
halo formation epoch by e.g.\@ \cite{Wechsler02}.

We specifically use the fitting formula given by 
\citet[][eq.\@ 6]{Wechsler06}:
\begin{equation}
b^{\rm ab}(M,c) = b_{\rm avg}(M)\times b_c(c|M/M_*) 
\label{eq:W06}
\end{equation}
where $b_{\rm avg}(M)$ is the mean halo bias at fixed mass,
$b_c(c|M/M_*)$ characterizes the concentration dependence of halo
bias, and $M_*$ is the characteristic mass of gravitational collapse
[quantitatively defined as $\sigma(M_*)=1.686$, where $\sigma(M)$ is
the r.m.s density fluctuation inside a sphere that encloses mass $M$].
The superscript ``ab'' refers to ``assembly bias,'' which we use as a
generic term for the dependence of halo bias on secondary variables,
based on the conjecture that these dependences arise through the
different formation histories of halos of the same mass.  We assume
this formula holds for all clusters included in our fiducial surveys,
although part of these clusters are outside the range where this
formula has been calibrated with simulations.  In addition, we note
that \cite{Wechsler06} calibrated this formula with $M_{\rm vir}$, while
the mass function and the halo bias we use are not always
well-calibrated with the same mass definition.  We ignore the
systematic errors that may be caused by these uncertainties.

\begin{figure*}[t!]
\plotone{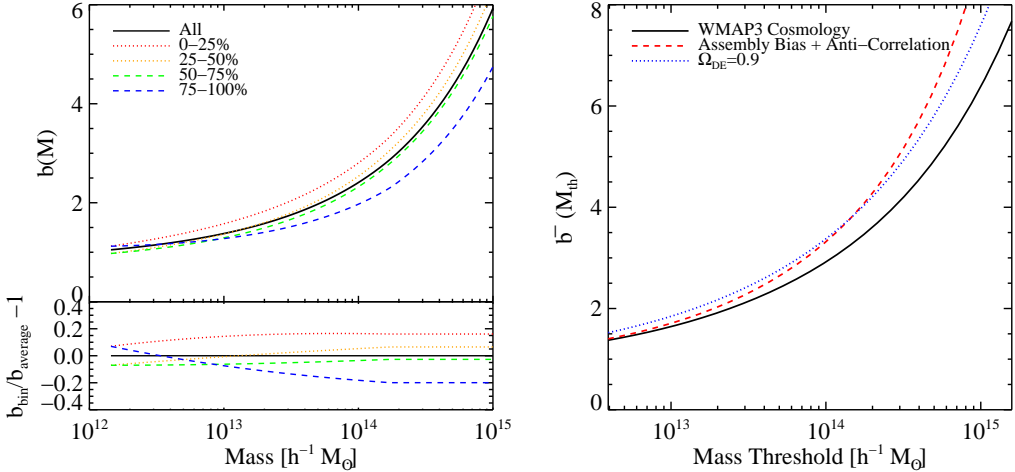}
\caption{ 
{\it Left:} Dependence of halo bias on concentration at
$z=0$ assumed in this work, based on the fitting formula of
$b^{\rm ab}(M,c)$ in \protect\cite{Wechsler06}.  We assume a WMAP3
cosmology and log-normally distributed concentrations at a given halo
mass.  Halos are binned by concentration into four quartiles, and the
halo bias of each quartile systematically deviates from the average
halo bias (\emph{solid curve}).  Above $10^{13.5} \hiMsun$, low concentration
halos (\emph{red and orange dotted curves}) are more clustered than high
concentration ones (\emph{green and blue dashed curves}) of the same mass.
The bottom panel shows the residual compared with the average halo
bias.  {\it Right panel:} Degeneracy between high dark energy density
and assembly bias.  The solid curve shows the cumulative bias
(eq.\@~[\ref{eq:cbias}], with the selection function nonzero above a
threshold $M_{\rm th}$) for the fiducial WMAP3 cosmology.  The dashed
curve shows the effect of assembly bias with the assumption of
perfectly anti-correlated cluster observable and concentration (see
\S\ref{sec:bias} and \ref{sec:assemblybias} for details).  This
correlation can mimic the effect of high dark energy density (here
assumed to be $\OmegaDE = 0.9$), shown as the dotted curve.
}
\label{fig:assemblybias}
\end{figure*}

The left panel of Figure~\ref{fig:assemblybias} illustrates how
concentration impacts halo bias in the fitting formula of
\citet[][]{Wechsler06}.  As can be seen, for $M\gtrsim10^{13.5}
\hiMsun$, low concentration halos are more clustered than high
concentration ones of the same mass.  This difference is potentially
significant: if the cluster observable is correlated with
concentration, one might measure cluster bias that differs from the
mean halo bias for random halos of the same mass.

The right panel of Figure~\ref{fig:assemblybias} shows how the effect
of assembly bias can resemble that of a high dark energy density, with
an extreme assumption of perfectly anti-correlated observable and
concentration.  Cumulative bias, which is relevant for halo samples
above a certain observable threshold (see eq.~[\ref{eq:cbias}]), is
plotted here.  As can be seen, if we tend to observe low concentration
halos, the effect of assembly bias (\emph{dashed curve}) makes the
observed halo bias higher than the mean halo bias (averaged over
random halos samples of the same mass) for the same fiducial cosmology
(\emph{solid curve}).  This effect mimics a high dark energy density
$\OmegaDE = 0.9$ (\emph{dotted curve}), since high $\OmegaDE$ will make
structures rarer and more clustered.  Thus, a wrong inference of
$\OmegaDE$ is possible if assembly bias is ignored in this case.  In
the following sections, we provide detailed formalism and analyses of
such systematics under the framework of the self-calibration of
observable--mass relation.


\section{Formalism}

\subsection{Counts-in-Cells Analysis and Basic Self-Calibration: A Review}\label{sec:self-calibration} 

In a pixelated galaxy cluster survey, halo bias is related to the
sample variance of cluster counts within the small sub-volumes of the
survey \citep{HuKravtsov03}.  Including the sample variance in a
counts-in-cells analysis allows one to ``self-calibrate'' the
observable--mass distribution, which is one of the main uncertainties
in modeling the surveys. This approach can thereby improve the dark
energy constraints relative to ``counts only'' experiments
\citep{Majumdar04,LimaHu04,LimaHu05,LimaHu07}.  In this section, we
review the basic self-calibration, closely following the formalism
developed by \cite{LimaHu04}.

Given a large-volume survey, consider a redshift slice which is
sufficiently thin to make evolution ignorable.  We then divide the
area of this slice into equal-area cells and count the clusters in
each cell.\footnote{ Here we suppress all redshift dependence in our
notation for simplicity.  In practice, we consider the redshift
dependence of the mass function, the halo bias, the observable--mass
distribution, and the comoving survey volume.  For readers of
\cite{LimaHu04}, note that our notation is slightly different.  Since
we consider a single redshift slice, our subscript $i$ indicates the
cell label of the same redshift, while in \cite{LimaHu04}, their
subscript $i$ indicates a cell of redshift $z_i$. } The number of
clusters in cell $i$, denoted by $N_i$, is affected by the Poisson
shot noise, which is modeled as $N_i \sim {\rm Poisson}(m_i)$.  This
Poisson mean $m_i$ varies from cell to cell due to the large-scale
clustering of matter and halos, and this fluctuation can be modeled as
a normal distribution $m_i \sim {\rm N}(\mbar, S)$, where $\mbar$ is
the mean halo abundance and $S$ is the sample variance.

In a given mass range, the mean number counts of clusters in cell $i$
depend on $\bar m$, the bias integrated over the mass range $\bar b$,
and the mass overdensity $\delta_i$ within this cell with respect to
the background:
\begin{equation}
m_i = \mbar (1+ \bar b\ \delta_i)\ .
\end{equation}
The sample variance then has the form
\begin{eqnarray}
S&=& \avg{(m_i-\mbar)^2} \nonumber\\
	&=& \mbar^2 \bar b^2 \sigma_{\rm V}^2 \ ,
\end{eqnarray}
where 
\begin{equation}
\sigma_{\rm V}^2 = \frac{1}{V^2}\int\frac{d^3\vec k}{(2 \pi)^3}W(\vec k)W^*(\vec k) P(k) \ .
\end{equation}
Here $P(k)$ is the matter power spectrum and $W(\vec k)$ is the
\emph{k}-space window function of a cell of volume $V$, normalized such that
$V = \int d^3\vec x W(\vec x)$.  Applying a counts-in-cells analysis,
$N_i$ of each cell can be measured, and $\mbar$ and $S$ can be
obtained from a likelihood analysis.  With additional knowledge of the
matter power spectrum, $\bar b$ can be obtained.

Note that this sample variance should be more rigorously defined as
the sample {\it covariance}
\begin{eqnarray}
S_{ij}&=& \avg{(m_i-\mbar)(m_j-\mbar)} \\
	&=& \mbar^2 \bar b^2 \sigma_{ij}^2 \ , \nonumber
\end{eqnarray}
with
\begin{equation}
\sigma_{ij}^2 = \frac{1}{V_i V_j}\int\frac{d^3\vec k}{(2 \pi)^3}W_i(\vec k)W^*_j(\vec k) P(k) \ . 
\end{equation}
In practice, our cell size is much larger than the correlation length
of clusters; thus, the correlations between different cells are
negligible. The off-diagonal elements are therefore much smaller then
the diagonal ones, and the matrix $S_{ij}$ reduces to a diagonal
matrix $S_{ij} = \delta_{ij} S$, whose dimension equals $n_c$, the
number of cells in the redshift slice.

We next relate these measurable quantities to theoretical models.  Let
$\Mobs$ denote the observed mass proxy (the observable) of galaxy
clusters.  Given a differential mass function $dn/dM$ and an
observable--mass distribution $P(\Mobs|M)$, the differential observed
cluster abundance is given as
\begin{equation}
\nMobs = \int dM\ \nM P(\Mobs|M) \ .
\end{equation}
In terms of the binning function $\phi(\Mobs)$---which is defined to
be equal to unity if $\Mobs$ falls in the bin corresponding to the
observable range, and zero otherwise---and the cell volume $V$, the
mean observed cluster abundance reads
\begin{equation}
\bar m = V \int d\Mobs\ \nMobs \phi(\Mobs) \ ,
\end{equation} 
which can be further simplified as
\begin{equation}
\bar m = V \int dM\ \nM \avg{\phi|M}
\label{eq:mean}
\end{equation}
if we define the selection function to be
\begin{equation}
\avg{\phi|M} = \int d\Mobs\ P(\Mobs|M) \phi(\Mobs) \ . 
\end{equation}
Given the halo bias $b(M)$, the bias integrated over the observable
bin similarly reads
\begin{equation}
\bar b = \frac{V}{\bar m}\int dM\ \nM b(M)  \avg{\phi|M} \ .
\label{eq:cbias}
\end{equation}
From equations~\ref{eq:mean} and~\ref{eq:cbias} we can see that if
both $\bar m$ and $\bar b$ are measured in the survey, the selection
function $\avg{\phi|M}$ can be self-calibrated.

In large-volume surveys, we often have several redshift bins and need
to consider how $\mbar$ and $S$ vary with redshift: $\mbar(z)$ and
$S(z)$.  The sample variance is then generalized to the matrix $\bold
S = {\rm diag} (\bold S_{ij}(z_1),\bold S_{ij}(z_2),...)$, where each
$\bold S_{ij}(z_k)$ has the dimension $n_c\times n_c$.  Similarly,
$\mbar$ is generalized as $\bold \mbar = (\bold\mbar(z_1),
\bold\mbar(z_2),...)$, with each $\bold\mbar(z_k) $ being a $n_c$
component vector.  For future reference, we further define $\bold M =
{\rm diag}(\bold\mbar)$ and $\C = \bold M +\bold S$; $\C$ is the
covariance matrix in the limit of large cluster numbers in a cell
($m_i\gg 1$; see \citealt{LimaHu04}).

Constraints on dark energy parameters are extracted from the
likelihood function that involves the counts-in-cells data, the
theoretical mean abundance, and the theoretical sample variance. For
theoretical forecasts, the Fisher matrix---the expectation value of
the second derivative of the minus log-likelihood function---is often
applied. For a combination of the Poisson shot noise and the Gaussian
sample variance, the Fisher matrix reads \citep{LimaHu04}
\begin{equation}
F_{\alpha \beta}=\bold{\bar m}^{T}_{,\alpha} \bold C^{-1} \bold{\bar m}_{,\beta} + \frac{1}{2} {\rm Tr}[\bold C^{-1} \bold S_{,\alpha}\bold C^{-1} \bold S_{,\beta}] 
\ ,
\label{eq:fisher}
\end{equation}
where the comma and subscript ${\alpha}$ indicates the partial
derivative with respect to model parameter $\theta_\alpha$.  The
Fisher matrix approach essentially approximates the likelihood
function as a Gaussian distribution near its maximum likelihood point,
and the curvature at this point is related to the constraints on the
model parameters.  The covariance matrix for model parameters is
approximated by the inverse of the Fisher matrix.  This basic picture
will play a key role in \S\ref{sec:fisher}, where we modify the Fisher
matrix formalism for assessing the systematic errors.


\subsection{Incorporating Assembly Bias into Self-Calibration}\label{sec:assemblybias}

We now incorporate assembly bias into the self-calibration formalism.
The formalism we outline below is relevant for any secondary parameter
which both affects the halo bias and correlates with the cluster mass
proxy. We specifically consider the secondary parameter to be the halo
concentration $c$ and refer to this dependence throughout as
``assembly bias.''  Note that although the halo concentration and
assembly history are generally expected to be tightly correlated
\citep{NFW97, Wechsler02}, they may not have exactly the same effect
on halo bias \citep[see e.g.][]{GaoWhite07}.

Let $b^{\rm ab}(M,c)$ be the halo assembly bias, which now depends on both
mass and concentration, and let $f(c|M)$ be the distribution of
concentrations for halos of mass $M$.  In this case, the
observable--mass distribution $P(\Mobs|M)$ needs to be generalized to
an observable--mass--concentration distribution $P(\Mobs|M,c)$.  With
the secondary parameter $c$, the mean abundance $\bar m$ takes the
form
\begin{equation}
\mbar = V \int dM\ \nM \int dc\ f(c|M) \avg{\phi|M,c} \ ,
\end{equation}
where
\begin{equation}
\avg{\phi|M,c} = \int d\Mobs\ P(\Mobs|M,c) \phi(\Mobs) \ .
\end{equation}
This mean abundance remains the same as equation~\ref{eq:mean} since
the concentration dependence only affects the halo bias but not the
mass function.  We thus require
\begin{equation}
\int dc\ f(c|M) \avg{\phi|M,c} = \avg{\phi|M} \ .
\label{eq:marginalization}
\end{equation}
On the other hand, the bias integrated over the observable range is
affected, and the analog of Equation \ref{eq:cbias} is
\begin{equation}
\bar b^{\rm ab} = \frac{V}{\mbar} \int dM\ \nM \int dc\  b^{\rm ab}(M,c) f(c|M) \avg{\phi|M,c} \ .
\end{equation}
The corresponding sample variance in this case reads
\begin{equation}
S_{ij}^{\rm ab} = \mbar^2 (\bar b^{\rm ab})^2 \sigma_{ij}^2 \ ,
\end{equation}
and we analogously define $\C^{\rm ab} = {\bold M} + {\bold S}^{\rm ab}$.
Replacing the corresponding matrices in Equation~\ref{eq:fisher}, we
obtain the Fisher matrix incorporating assembly bias.

\begin{figure}[t!]
\plotone{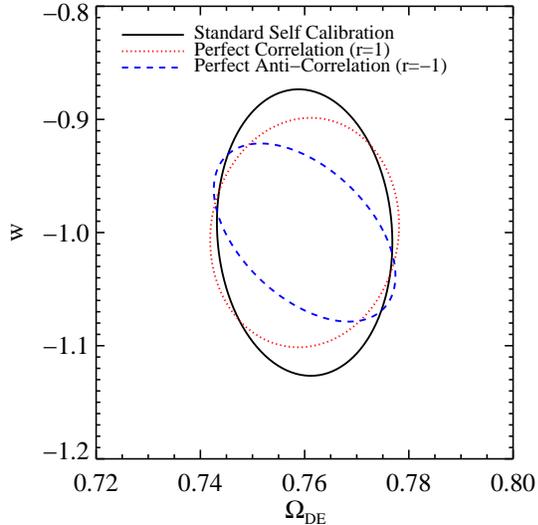}
\caption{
Improvement of dark energy constraints assuming a thorough modeling of
assembly bias and knowledge of the cross-correlation relating $\Mobs$
---the cluster's mass estimate based on a cluster observable---and
$c$, the halo's concentration parameter.  All error ellipses include
the $68\%$ confidence regions in the $\OmegaDE$--$w$ plane.  The solid
ellipse shows the fiducial model of zero observable--concentration
correlation ($r=0$), in which case assembly bias has no effect.  The
dotted/dashed ellipse corresponds to an observable which is perfectly
correlated/anti-correlated with concentration ($r=1$/$-1$). If
assembly bias is correctly modeled, the sensitivity of assembly bias
to $M_*$ slightly improves dark energy constraints.
}
\label{fig:EllipseAB}
\end{figure}

The difference between $P(\Mobs|M,c)$ and $P(\Mobs|M)$ depends on how
$\Mobs$ correlates with $c$.  We leave these details to
\S\ref{sec:correlation} and simply state here that our parametrization
depends on the cross-correlation coefficient $r$ relating $\Mobs$ and
$c$ at fixed halo mass.  When $r=0$, assembly bias has no impact on
self-calibration; when $r=\pm 1$, the impact of assembly bias is
maximized.  Figure~\ref{fig:EllipseAB} demonstrates the formalism
described above (with an SPT survey assumption and a WMAP3 cosmology,
see \S\ref{sec:implementation}) and shows how the correlation between
$\Mobs$ and $c$ changes the constraints on dark energy parameters,
assuming that we have thorough knowledge of assembly bias and that
$r=\pm1$ (\emph{dotted and dashed curves}).  As can be seen, correlation
between $\Mobs$ and $c$ actually improves the dark energy constraints
if $r$ is known a priori.  This improvement is presumably due the
dependence of bias on $M_*$, which is also sensitive to dark energy,
although we also note that the assumption of self-similarity in
$M/M_*$ needs to be assessed in the dark energy-dominated regime.  In
addition, with the knowledge of $r$, the scatter in $\Mobs$ actually
contains the information of halo concentration, which may also improve
cosmological constraints.  These extreme cases are mainly for
demonstration, since we are unlikely to have sufficient astrophysical
knowledge to specify both the assembly bias and this correlation.
However, if individual concentrations can be measured for the most
massive clusters (where the impact of assembly bias is most severe),
they could provide observational evidence of assembly bias and
increase the efficacy of self-calibration.

In the following sections, we explore the question: if one were to
perform the self-calibration analysis ignoring the effects of assembly
bias (effectively, assuming $r=0$), how would the estimated
cosmological parameters be biased?  As we shall see, the answer
sensitively depends on $r$ and on the scatter in the observable--mass
distribution.  We next include $r$ as a free parameter in the Fisher
matrix analysis and consider the effect of marginalization over $r$.
However, a caveat for applying the Fisher matrix here is that since
$r$ is bound to the range $[-1,1]$, the likelihood function for $r$
may not be well-approximated as Gaussian if $r$ is close to $\pm 1$.
Because the Fisher matrix is based on this Gaussian approximation, it
may not apply to the case when $r$ approaches $\pm1$.  On the other
hand, our fiducial choices of this parameter, which are in the range
$|r| \leq 0.5$, may circumvent this problem.

\subsection{Biased Parameter Estimation from Ignored Systematics: A Modified Fisher Matrix Formalism}\label{sec:fisher}

In \S\ref{sec:bias}, we described how ignoring the impact of assembly
bias can potentially lead to biased cosmological parameter estimates.
In this section, we modify Fisher matrix formalism to quantitatively
assess the significance of this systematic.  We focus on how the
parameter estimates are biased due to a wrong model assumption, and
how significant this systematic error is when compared with
statistical uncertainties.  This formalism is motivated by the
standard Fisher matrix formalism as presented in \cite{Tegmark97KL}.

We generally consider two models, denoted by model $A$ and model $B$,
each of which describes a data set $\vec x$ based on a parameter
$\theta$.  Here $\theta$ can be generalized to a vector denoting a set
of parameters ($\theta_i$ values).  We assume that the observed data set
$\vec x$ is well described by model $B$ but is mistakenly analyzed
according to model $A$.  If $\theta_t$ denotes the true parameter in
model $B$ that corresponds to the observed data set $\vec x$, we are
interested in how the estimated parameter $\hat \theta$ recovered
based on model $A$ differs from $\theta_t$. Our quantitative analysis
can be summarized as follows:
\begin{enumerate}
\item Our starting point is the likelihood function $L_A(\vec
x|\theta)$ for model $A$.  The data set $\vec x$ is assumed to be
drawn from the probability distribution $P_B(\vec x|\theta_t)$ for
model $B$; in order to relate $\theta$ to $\theta_t$, we take average
over $\vec x$ to compute $\avg{\ln L_A(\theta)|\theta_t}$.
\item We take the point $\hat\theta$ which maximizes $\avg{\ln
L_A(\theta)|\theta_t}$ as our estimator for the recovered cosmology.
This step defines the function $\hat\theta(\theta_t)$, the recovered
model parameter varying with the input parameter $\theta_t$.  We are
particularly interested in $\delta\theta =
\hat\theta(\theta_t)-\theta_t$, which is the systematic error in
parameter inference due to assuming an incorrect model.\footnote{An
alternative approach is to first use $L_A(\vec x|\theta)$ to compute
the maximum likelihood estimator $\hat\theta(\vec x)$.  Since
$\hat\theta$ is now a function of the data $\vec x$, one could use
$P_B(\vec x|\theta_t)$ to compute the expectation value
$\avg{\hat\theta|\theta_t}$.  However, this approach is not
analytically tractable.  }
\item In order to assess the significance of the systematic error
$\delta\theta$, we compare it against the statistical uncertainty in
$\theta$.  We calculate the modified Fisher matrix $ \bold{\tilde
F}_{ij}(\theta_t) =\avg{\partial^2(-\ln L_A)/\partial\theta_i\partial
\theta_j| \theta_t}$ and obtain the corresponding error bar
$\sigma_{\theta_i}^2= {(\bold{\tilde F}^{-1})_{ii}}$.  The systematic
error is significant if $\delta\theta_{i} \gtrsim \sigma_{\theta_i}$.
\end{enumerate}
A detailed derivation when both $P_A(\vec x|\theta)$ and $P_B(\vec
x|\theta)$ are Gaussian can be found in the Appendix.

In this study, model $A$ represents the standard self-calibration
analysis that ignores assembly bias, while model $B$ is
self-calibration analysis that includes the impact of assembly bias.
The data set $\vec x$ is the number counts in each of the cells under
consideration.  The systematic errors of the recovered parameters are
given by
\begin{equation}
\delta\theta_j =\sum_i (\bold{F}^{-1})_{ij} {\rm Tr}\{\frac{1}{2}{\C^{-1}\C_{,i}\C^{-1}(\C^{\rm ab}-\C)}\} \ ,
\end{equation}
where $\C^{\rm ab}$ is the covariance matrices with assembly bias, and
$\C$ and ${\bold F}$ are the same as those in
equation~\ref{eq:fisher}.  The modified Fisher matrix reads
\begin{equation}
{\tilde F}_{ij}={\bold{\bar m}}^{T}_{,i} \bold C^{-1} {\bold{\bar m}}_{,j} + \frac{1}{2} {\rm Tr}[\bold C^{-1} \bold S_{,i}\bold C^{-1} \bold S_{,j}\C^{-1}\C^{\rm ab}]  \ ,
\end{equation}
in which the modification comes from the change of covariance matrix
due to assembly bias (see the Appendix).
We note that similar formalisms arising from different approaches can
be found in e.g.\@ \citet{Knox98}, \citet{Huterer01},
\citet{HutererLinder07}, and \citet{Amara07}.

\begin{figure*}[t!]
\plotone{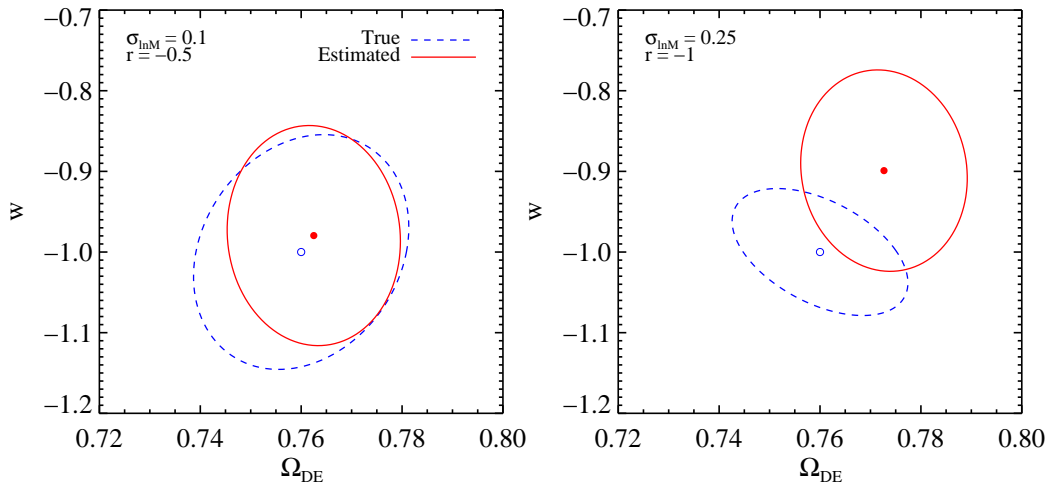}
\caption{
Systematic errors due to ignoring existent assembly bias.  Here we
assume two sets of scatter and correlation values and perform the
analysis discussed in \S\ref{sec:fisher}, with an SPT survey
assumption and a WMAP3 cosmology (see \S\ref{sec:implementation}).
The open circles and dashed ellipses show the true parameter values
and the $68\%$ confidence regions with assembly bias correctly
included.  The solid circles and the solid ellipses show the estimated
values and $68\%$ confidence regions if assembly bias is completely
ignored.  The left panel shows that for a moderate assumption of
$\sigma_{\rm\ln M }=0.1$ and $r=-0.5$, the systematic errors are
$0.22$ and $0.23\sigma$ for $\OmegaDE$ and $w$, respectively; in
this case the effects of assembly bias are ignorable. On the other
hand, the right panel shows that for an extreme assumption of
$\sigma_{\rm\ln M }=0.25$ and $r=-1$, the systematic errors are
$1.14$ and $1.2\sigma$ for $\OmegaDE$ and $w$, respectively; in
this case the effects of assembly bias are significant.
}
\label{fig:estimator}
\end{figure*}

Figure \ref{fig:estimator} illustrates the results of our formalism as
applied to the self-calibration analysis for an SPT-like survey in the
specified WMAP3 cosmology (see \S\ref{sec:implementation} for details
of implementation and assumptions).  In each panel, the open circles
indicate the assumed true values, while the filled circles show the
recovered parameters from a self-calibration analysis that ignores
assembly bias.  The ellipses include the $68\%$ confidence regions in
the $\OmegaDE$--$w$ plane; the dashed ellipses correspond to
correctly-modeled assembly bias (assuming that we know the correlation
coefficient $r$ a priori; $r$ will be mathematically defined in
\S\ref{sec:correlation}), while the solid ellipses correspond to the
ignored assembly bias. Note that the shape of the confidence regions
can also be changed by this systematic.  The left panel shows the
assumption of a small $\Mobs$--$M$ scatter and low $\Mobs$--$c$
correlation ($\sigma_{\rm\ln M }=0.1$ and $r=-0.5$), and the systematic
errors are $0.22$ and $0.23\sigma$ for $\OmegaDE$ and $w$,
respectively; the deviations of the parameter estimates are much less
than the statistical uncertainties.  The right panel shows the
assumption of a larger scatter and perfectly anti-correlated $\Mobs$
and $c$ ($\sigma_{\rm\ln M }=0.25$ and $r=-1$), and the resulting
systematic errors are $1.14$ and $1.2\sigma$ for $\OmegaDE$ and
$w$, respectively; these deviations are significant and cannot be
ignored.  We thus expect the impact of assembly bias will be stronger
if the observable--mass relation has a large scatter and if $\Mobs$ is
strongly correlated with $c$.  The exact dependence of systematic
error on these two quantities will be fully explored in \S
\ref{sec:results}.


\section{Implementation}\label{sec:implementation}

\subsection{Parameterizing the Observable-Concentration Correlation}\label{sec:correlation}

In the absence of assembly bias, we follow \citet{LimaHu05} to
parameterize the observable--mass relation $P(\ln\Mobs|M)$. Given halo
mass $M$, the corresponding log observables $\ln\Mobs$ are modeled as
a Gaussian distribution with mean $\ln M + \ln\Mbias$---where $\Mbias$
specifies the offset between the estimate mass and the true mass---and
variance $\sigmalnM^2$.  This parameterization serves as the standard
case as we generalize $P(\ln\Mobs|M)$ to $P(\ln\Mobs|M,c)$ for
analyzing the effect of assembly bias.

A priori, we do not know exactly how the estimated mass of a cluster
$\Mobs$ will depend on the cluster's concentration $c$, that is, the
correct parameterization for $P(\ln\Mobs|M,c)$.  In detail, this
relation may depend on both physical and observational
effects. However, we would like to demand a simple wish-list of
properties of our parameterization:
\begin{enumerate}
\item When marginalized over concentration, $P(\ln\Mobs|M,c)$ should
reduce to the Gaussian distribution $P(\ln\Mobs|M)$ of the fiducial
case (as required by eq.~[\ref{eq:marginalization}]), independent
of any new parameters introduced (i.e.\@ we should keep the total
$\ln\Mobs$--$ \ln M$ scatter fixed).
\item In order to study how self-calibration is affected as the
dependence of $\Mobs$ on $c$ is ``turned on,'' the parameterization
should have a tunable parameter. When this tunable parameter is set to
zero, our analysis should reduce to the standard case.
\end{enumerate}
In the interest of simplicity, we take $P(\ln\Mobs|M,c)$ to be
Gaussian in $\ln\Mobs$, and assume that the halo concentration
slightly shifts $\ln\Mobs$ relative to $\ln M$, so that the mean and
the variance of $\ln\Mobs$ are given by
\begin{eqnarray}
\avg{\ln \Mobs|M,\tc} &=& \ln M + \ln M_{\rm bias} + r\sigmalnM\tc \\
\Var(\ln \Mobs|M,c) &=& \sigmalnM^2(1-r^2) \ .
\end{eqnarray}
In the above expressions, $r$ is the correlation coefficient between
$\ln\Mobs$ and $\tc$ at fixed $\ln M$, $\sigmalnM$ is the scatter in
$\ln\Mobs$ at fixed $M$, and $\tc$ is defined via
\begin{equation}
\tc = \frac{\ln c-\avg{\ln c|M}}{\sqrt{\Var(\ln c|M)}} \ .
\end{equation}
Note that when $r=0$, all of the observed scatter in $\ln\Mobs$ at
fixed $\ln M$ is intrinsic, and our model reduces to the standard
case.  Conversely, for $r=1$, the scatter in $\ln\Mobs$ at fixed $\ln
M$ is entirely due to the scatter in halo concentration at fixed mass.
As a consistency check, we find that if we marginalize $P(\ln
\Mobs|M,c)$ over concentration (assuming a log-normal distribution for
$c$ at fixed mass, see e.g.\@ \citealt{Jing00}, \citealt{Bullock01},
and \citealt{Neto07}), the resulting distribution $P(\ln\Mobs|M)$ is
exactly that of the standard case; that is, our parameterization
preserves the total scatter in $\ln\Mobs$ at a given $\ln M$.


\subsection{Survey Assumptions, Cosmological Models, and Nuisance Parameters}

\begin{deluxetable}{ccccccc}
\tabletypesize{\scriptsize}
\tablecaption{Survey Assumptions\label{tab:surveys}}
\tablewidth{0pt}
\tablehead{
    \colhead{Survey} &
    \colhead{$M_{\rm th}$} &
    \colhead{Bin Size} &
    \colhead{$N_{\rm bins}$} &
    \colhead{Area} &
    \colhead{$z_{\rm max}$}\\ 
    \colhead{} & 
     \colhead{$(\hiMsun)$} & 
    \colhead{$(\Delta {\rm log}_{10}\Mobs)$} &   
    \colhead{} &
    \colhead{$(\rm{deg^2})$}
}    
\startdata
SDSS (optical)  & $10^{13.5}$ & 0.5 & 3   & 7500   & 0.3  \\  
DES (optical)    & $10^{13.5}$ & 0.5 & 3   & 5000   & 1     \\
SPT (SZ)          & $10^{14.2}$ & 1    & 1  & 4000   & 2     \\
LSST (optical)  & $10^{13.5}$ & 0.5 & 3   & 20000   & 2  \\   
\enddata
\tablecomments{All surveys use cells of area $10\deg^2$ and $\Delta z = 0.1$}
\end{deluxetable}

With the Fisher matrix analysis, we statistically forecast the
systematic effects for four galaxy cluster surveys: the Sloan Digital
Sky Survey (SDSS, \citealt{SDSS}; assuming the volume using
photometric data), the Dark Energy Survey
(DES\footnote{See http://www.darkenergysurvey.org/}), the South Pole
Telescope (SPT\footnote{See http://pole.uchicago.edu/}), and the Large
Synoptic Survey Telescope (LSST\footnote{See http://www.lsst.org/}).  The
survey areas are assumed to be $7500\ \deg^2$ for SDSS, $5000\ \deg^2$
for DES, $4000\ \deg^2$ for SPT, and $20000\ \deg^2$ for LSST, with
survey depths of $z_{\rm max}=0.3, 1.0, 2.0$ and $2.0$ respectively.  The
cells used for the counts-in-cells analysis are assumed to have an
area $10\ \deg^2$ and redshift interval $\Delta z =0.1$.  We assume
clusters with $\Mobs \geq10^{14.2}\ \hiMsun$ are observed by SPT, and
perform no mass binning.  For SDSS, DES, and LSST, the observational
threshold is assumed to be $\Mobs\geq 10^{13.5}\ \hiMsun$, and the
counts in each of these surveys are binned in three observable bins.
The survey parameters for all four surveys are detailed in
Table~\ref{tab:surveys}.

While the mass threshold of SZ observations has little redshift
dependence \citep[e.g.][]{Carlstrom02}, the mass threshold of optical
surveys has more uncertainties.  Clusters with mass $10^{13.5}\
\hiMsun$ have been shown to be detectable, with high purity and
completeness, with more than 10 bright red galaxies ($\sim 0.4 L_*$)
in the SDSS photometric survey out to $z \sim 0.3$ \citep{Koester07,
Johnston07}.  We note that our choice of the minimum mass for the
optical surveys assumes that such clusters can still be detected with
high purity and completeness out to the maximum redshift $z_{\rm max}$.
This assumption may be reasonable out to $z=1$, where clusters have
been shown to have a robust red sequence, but the efficacy of this
method will eventually break down at higher redshifts.  In any case,
it will need to be tested in detail with both realistic simulations
and the data itself.  We note that for LSST, one may wish to detect
clusters using peaks in the lensing shear instead of from assumptions
about the galaxy distribution \citep[e.g.][]{Kaiser95,Hennawi05}, in
which case self-calibration could serve as a consistency check for the
predictions for the observed shear signal made directly from
simulations.  In \S\ref{sec:results}, we consider one example case for
LSST, which has similar assumptions to the lower $z_{\rm max}$ optical
surveys, for reference.

\begin{deluxetable*}{ccccccc}
\tabletypesize{\scriptsize}
\tablecaption{Fiducial Cosmologies\label{tab:cosmology}}
\tablewidth{0pt}
\tablehead{
    \colhead{Cosmology} &
    \colhead{$\OmegaDE$} &
    \colhead{$w$} &
    \colhead{$\delta_\zeta(k=0.05{\rm Mpc^{-1}})$} &
    \colhead{$n$} &
    \colhead{$\Omega_bh^2$} &
    \colhead{$\Omega_mh^2$} 
}
\startdata
WMAP1 & 0.73 & -1 & $5.07\times10^{-5}$ & 1        & 0.024   & 0.14\\ 
WMAP3 & 0.76 & -1 & $4.53\times10^{-5}$ & 0.958 & 0.0223 & 0.128\\  
\enddata
\tablecomments{All of our forecasts assume Plank-like priors: $\sigma(\ln \zeta)=\sigma(\ln \Omega_mh^2) = \sigma(\ln \Omega_bh^2) = \sigma(n) =0.01$, except for $\OmegaDE$ and $w$.}
\end{deluxetable*}

In this work, we consider two sets of cosmological parameters, namely
the best fit cosmologies to WMAP1 \citep{WMAP1} and WMAP3
\citep{WMAP3}, whose parameter values are listed in
Table~\ref{tab:cosmology}.  Both of them are flat $\Lambda$CDM
cosmologies but differ mainly in the relative contribution of dark
energy to the global energy density, in the normalization of
fluctuations ($\delta_\zeta$ or $\sigma_8$), and in the spectral index
($n$).  The impact of these differences on our analysis will be
presented in \S\ref{sec:results}.  In our statistical forecast, we do
not put any priors on dark energy parameters, but we assume
Planck-like priors on the rest of the cosmological parameters (see
Table~\ref{tab:cosmology}).  Finally, in our forecast models we use
the halo mass function by \citet{Jenkins01}, the bias function by
\citet{Sheth01}, and the assembly bias $b^{\rm ab}(M,c)$ found by
\citet[][this assumption was shown in Fig.\@ 1]{Wechsler06}.

With regard to the observable--mass relation, our model involves three
nuisance parameters: the bias in the estimated mass ($\ln M_{\rm bias}$),
the scatter of $\ln\Mobs$ given $\ln M$ ($\sigma_{\rm \ln M}$), and the
cross-correlation coefficient between $\ln\Mobs$ and the normalized
halo concentration $\tc$ ($r$).  Throughout, we take $\ln M_{bias}=0$
as our fiducial model.  Our choice for the fiducial values for the
scatter and the cross-correlation coefficient in each of the surveys
requires further discussion.

Let us first focus on the scatter. For an SPT-like survey, the
observational mass proxy is the SZ decrement of the cosmic microwave
background due to the hot, ionized gas permeating the inter-cluster
medium.  At present, this scatter has only been predicted from
numerical simulations but has not been determined from observations.
\cite{WhiteM02} argued that the three main sources of scatter are the
evolution of the $M$--$T$ relation, asphericity in the matter
distribution, and line-of-sight projection.  \cite{Motl05} and
\cite{Nagai06} showed that the scatter is 10\%-15\%, and the scaling
relation is insensitive to the detailed physical processes involved in
galaxy formation, with a good agreement with self-similar models.
However, \cite{Shaw07} showed that at least 20\% intrinsic scatter
exists due to the internal properties of galaxy clusters.  They also
demonstrated that this scatter could be reduced by choosing different
aperture radius for defining $M$ and $Y$, or by removing cluster samples
with many substructures.  Moreover, it may be possible to reduce the
scatter even further using cluster structural properties.  For
example, \cite{Afshordi07} proposed a ``fundamental plane'' among the
cluster mass, the total SZ flux, and the SZ half-light radius
$R_{\rm SZ,2}$; in simulations, this relation reduced the scatter in mass
estimates to $\sim$ 14\%.  Further, \cite{Haugboelle07} found that by
constructing an empirical model for the SZ profile, which includes a
scaling parameter $r_0$, they could reduce the scatter down to 4\%.
In this work, we take the largest of these range of values, namely
20\%, as our fiducial scatter for SPT.  If SPT is insensitive to halo
assembly bias for this largest possible scatter, then it will also be
insensitive for smaller values of scatter.

In optical surveys, the usual observational mass proxy is the optical
richness, namely the galaxy number in a galaxy cluster. Other choices
are also possible, including the total optical luminosity or
combinations of parameters \citep[e.g.][]{Rykoff07,Becker07,Reyes08}.
Determining a reasonable choice for the scatter for a DES-like survey
is somewhat less straightforward, as predictions from simulations are
less robust and the scatter can highly depend on both the richness
measure and the cluster finder.  \cite{Gladders07} applied a
self-calibration analysis to a catalog from the Red-Sequence Cluster
Surveys (RCS), finding the fractional scatter $f_{\rm sc}$ to be $0.69 \pm
0.20$ and $0.71^{+0.19}_{-0.17}$ (based on different priors).  Using
the velocity information of galaxies in maxBCG clusters,
\cite{Becker07} estimated that the optical richness had a mass
dependent scatter which varied from about $0.75$ for massive clusters
to $1.2$ for group scale objects.  Cross-correlation with the X-ray
data on these same clusters suggests a considerably smaller scatter of
about $0.5$ (E.\@ Rozo et al.\@ in preparation).  Here, we choose $0.5$ as
our fiducial scatter value for two reasons: First, as we shall see,
even with this amount of scatter, halo assembly bias has a significant
impact on self-calibration studies; this small scatter thus provides a
baseline value for the impact of assembly bias.  Second, we note that
current optical richness estimates have all used fairly crude measures
of richness.  We think it is highly probable that in the near future
we will start seeing richness measures that are considerably more
strongly correlated with mass than those used at present.  Thus, we
have opted to select a scatter value that is closer to the lowest
scatter estimated in current samples.

We now turn to the correlation relating observable and concentration,
$r$.  Currently, there are no observational constraints on $r$ for
either optical or SZ mass proxies.  We also find no quoted values for
the correlation between SZ and halo concentration in the literature,
although we note that \citet[][]{Reid06} and \citet[][]{Shaw07}
investigated the impact of halo concentration on the scatter in
$Y_{\rm SZ}$ and found that a considerable fraction of the scatter in SZ
is due to variations in halo concentration at fixed mass.  In this
work, we choose the fiducial value to be $r=0.4$, which is the value
observed in simulations of clusters using the hydrodynamical ART code
(D.\@ Rudd 2008, private communication).

The $\Mobs$--$c$ correlation $r$ for optical clusters is somewhat
better understood.  We are not aware of current observational
constraints on the correlation between optical cluster richness and
halo concentrations, although with a sufficiently large sample, this
value could in principle be measured from lensing data.
\citet[][]{Zentner05} and \citet[][]{Wechsler06} have shown that the
amount of substructures in a cluster-size halo is negatively
correlated with halo concentration.  On the other hand, selection
effects could modify this correlation.  For instance, high
concentration halos might, on average, be assigned higher richness
than low concentration halos of the same mass due to the larger galaxy
density near the cluster core.  In this work, we choose $r=-0.5$ as
our fiducial value, which is roughly consistent with the numerical
results of \citet[][]{Zentner05} and \citet[][]{Wechsler06}.  We note
that the results presented here assume that all of these parameters
are constant with both redshift and mass.


\begin{figure*}[t!]
\plotone{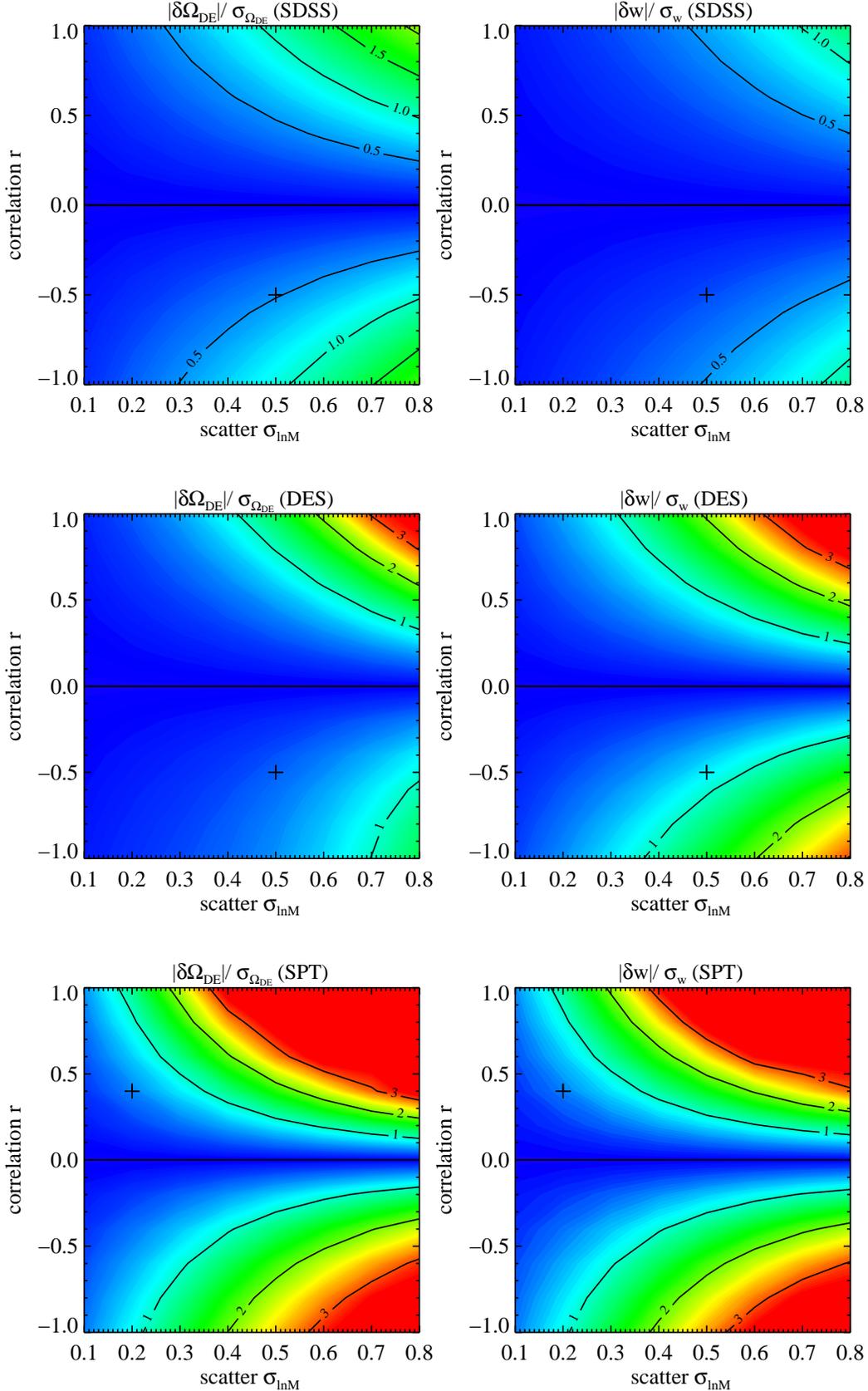}
\caption{
Systematic errors for $\OmegaDE$ (\emph left) and $w$ (\emph right)
estimators, as a function of scatter in the observable given mass
($\sigmalnM$) and the correlation between observable and halo
concentration ($r$).  The ratios of the systematic error and the
statistical uncertainty ($|\delta\theta|/\sigma_{\rm\theta}$) are shown for
three of our main survey assumptions: SDSS, DES, and SPT, from top to
bottom.  High scatter value and strong correlation/anti-correlation
correspond to high deviation of estimators.  We also mark the fiducial
values of $\sigmalnM$ and $r$ in each panel according to our current
knowledge from observations and numerical simulations. We note that
these plots are also applicable to other surveys (e.g. X-ray) for
which the survey volume and mass threshold are the same as those
assumed here.  See \S\ref{sec:results} for discussion.
}
\label{fig:Contours}
\end{figure*}

\section{Results and Discussion}\label{sec:results}

We now present the effect of assembly bias for a set of specific
assumptions about galaxy cluster surveys.  We focus on the systematic
errors in the two dark energy parameters $\OmegaDE$ and $w$ when
compared with the statistical errors expected from each survey,
$|\delta \OmegaDE|/{\sigma_{\Omega_{\rm DE}}}$ and $|\delta
w|/\sigma_w$.  Figure~\ref{fig:Contours} shows how these two ratios
vary with the scatter in $\ln\Mobs$--$\ln M$ ($\sigma_{\rm \ln M}$) and
the cross-correlations coefficient of $\ln\Mobs$--$\tc$ ($r$) for our
fiducial SDSS, DES, and SPT surveys.  All plots assume a WMAP3
cosmology.  As can be seen, a high degree of correlation and/or large
scatter can result in significantly biased cosmological estimates for
both DES and SPT, while for the current SDSS the statistical
uncertainty is sufficiently large that halo assembly bias is
insignificant.

How halo assembly bias differently affects DES and SPT is worth
discussing.  For a fixed scatter and correlation coefficient, the
cosmological constraints from DES are considerably less biased than
those of SPT.  The reason for this difference is two-fold. First, DES
clusters probe a lower mass scale than SPT clusters do.  This
difference is important because the effect of concentration on halo
bias is important for high mass halos, but non-existent for halos of
mass near $M_*\sim 10^{13} \hiMsun$ \citep{Wechsler06}.  Consequently,
the cosmological constraints coming from low mass clusters (groups)
should be unbiased.\footnote{Although the low mass clusters (groups)
are less affected by assembly bias, they are subjected to more
statistical errors.  For the most constraining power, the choice of
mass threshold should be made based on the scatter, the completeness,
and the purity.}  The second important difference between SPT and DES
is that in our fiducial surveys we have assumed binned counts for DES
clusters but only thresholded counts for SPT clusters.  Consequently,
all the cosmological information provided by the shape of the halo
mass function (which is unaffected by assembly bias) does not
contribute to the SPT constraints.  Thus, SPT constraints are
considerably more sensitive to the effects of assembly bias than the
DES constraints given the same scatter and correlation coefficient.

That is not, however, the end of the story.  In order to fairly
compare SPT to DES, one also needs to consider the regions of
parameter space relevant to each of these surveys.  We noted earlier
that numerical simulations predict that the intrinsic scatter in the
SZ signal is approximately $20\%$ or even less
\citep[e.g.][]{Motl05,Nagai06,Shaw07,Haugboelle07}.  As can be seen
from the bottom panels of Figure~\ref{fig:Contours}, these scatter
values do not result in significant biasing of the recovered
cosmological parameters for any value of $r$.  Although the expected
intrinsic scatter is small, the projection effect may raise or even
dominate the total scatter \citep[see e.g.][]{WhiteM02,Hallman07}.
\citet{Holder07} found that the SZ background can generate errors
larger than $20\%$ in recovered flux if $\sigma_8$ is near 0.7.  If
the extra scatter is due to the randomly aligned structures along the
line of sight, then we do not expect assembly bias to have a significant
impact. On the other hand, if the projection effect is dominated by
nearby structures, the extra scatter due to projection will be
strongly correlated with the environment, resulting in higher
correlation between concentration and observable.  In fact, if the
scatter due to projection is as high as \citet{Holder07} predicted and
is also dominated by nearby correlated structure, the effect of
assembly bias may be strengthened.

Photometric surveys like DES, in contrast, are very likely to be
sensitive to the impact of assembly bias.  In this case, we know that
the optical richness--mass relation has a scatter $\gtrsim 50\%$
(e.g. \citealt{Gladders07, Becker07}; E.\@Rozo et al.\@ in preparation).
As can be seen in the middle panels of Figure~\ref{fig:Contours}, even
moderate correlations between $\Mobs$ and $c$, e.g.\@ $|r|\gtrsim
0.5$, can result in significant biasing of the recovered cosmological
parameters.  It is likely, therefore, that cosmological analysis of
the DES optical cluster sample will need to include halo assembly bias
in order to avoid systematic errors in dark energy inference, unless
the analysis can be done with an observable that is more tightly
correlated with mass.  On the other hand, DES will have additional
mass measurements, including its weak lensing and SZ signals from SPT.
If these measurements are included in the analysis, the effect of
assembly bias may be diminished.  Observational inference of assembly
bias may even be possible with mass profile measurements.

We note that in these figures, the only differences are between the
survey volumes and the threshold in $M_{\rm obs}$, without assuming any
other information specific to the optical or SZ surveys.  Therefore,
these results are applicable to other surveys (e.g.\@ X-ray surveys)
with the same survey conditions.

\begin{figure}[t!]
\plotone{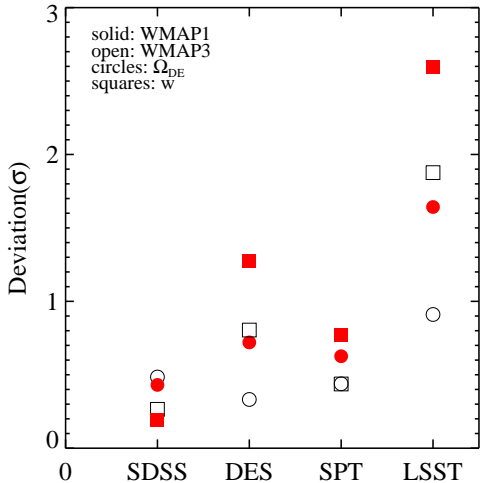}
\caption{
Impact of assembly bias for two different cosmologies and four
survey conditions.  The ratio $|\delta\theta|/\sigma_\theta$ for
$\OmegaDE$ and $w$ are plotted as circles and squares respectively.
Solid and open symbols are for WMAP1 and WMAP3 cosmologies
respectively.  DES and LSST are clearly sensitive to assembly bias,
while SPT is marginally sensitive to it, with the effect being
stronger for WMAP1 than WMAP3.  A current SDSS-like survey is not
sensitive to assembly bias.  (Fiducial values assumed for other
parameters include: $r=-0.5$ and $\sigma_{\rm \ln M}=0.5$ for SDSS, DES,
and LSST, and $r=0.4$ and $\sigma_{\rm \ln M}=0.2$ for SPT)
}
\label{fig:surveys}
\end{figure}

In Figure~\ref{fig:surveys} we (1) explore the systematic effects due
to assembly bias under different cosmological parameters (WMAP1 and
WMAP3) and (2) extend the calculation to include an assumption for an
LSST-like optical cluster survey.  (For SDSS, DES, LSST: $r=-0.5$ and
$\sigma_{\rm\ln M}=0.5$. For SPT: $r=0.4$ and $\sigma_{\rm\ln M}=0.2$.)
First, the most relevant difference between WMAP1 and WMAP3 is that
WMAP3 has higher $\OmegaDE$ and lower normalization $\delta_\zeta$ or
$\sigma_8$ values, as listed in Table~\ref{tab:cosmology}.  As a
result, the WMAP3 cosmology has fewer clusters, and the sample
variance of the clusters is smaller. These differences increase the
statistical errors of the surveys \citep[see also][]{LimaHu07}, thus
making the impact of assembly bias less significant in the WMAP3
cosmology than in the WMAP1 cosmology.  Overall, our main conclusions
remain unchanged.  Second, for LSST, we find the systematic due to
assembly bias is very significant for our fiducial values; this
systematic is likely to be significant for even small values of
$\sigmalnM$ and $r$.

We especially note that the systematic of assembly bias impacts
$\OmegaDE$ and $w$ differently.  From the survey point of view,
increasing $z_{\rm max}$ above $1$ largely improves the constraints on
$w$, but barely improves the constraints on $\OmegaDE$.  For $w$, the
systematic error due to assembly bias increases monotonically with
$z_{\rm max}$, while for $\OmegaDE$, this systematic error somewhat
cancels and then changes its sign as $z_{\rm max}$ increases.  This
difference is due to the fact that $\OmegaDE$ and $w$ affect the
observed large-scale structure differently in different regimes.  Dark
energy affects the observed large-scale structure through two
mechanisms: the growth function and the comoving volume.  High
$\OmegaDE$ and high $w$ both result in stronger suppression of
structure.  On the other hand, the volume dependence works
differently: high $\OmegaDE$ and low $w$ correspond to larger volumes.
The effects work in opposite directions for $\OmegaDE$; higher
$\OmegaDE$ leads to less structure but more volume. Before the onset
of dark energy domination, the comoving volume effect dominates; after
dark energy take-over, the growth function effect dominates.  Thus
near the onset of dark energy domination, the observed structure is
insensitive to $\OmegaDE$, leading to no extra information from this
regime.  The effects of assembly bias on $\OmegaDE$ before and after
the dark energy domination have opposite signs and thus cancel each
other.  On the other hand, for $w$, both effects work in the same
direction; thus, including more survey volume will always increase the
amount of information on $w$, and the systematic effects do not
cancel.  That is why the systematic effect of assembly bias on $w$
increases monotonically with $z_{\rm max}$.

Another interesting question is how the constraints on cosmological
parameters are degraded if we include $r$ as an additional nuisance
parameter that needs to be marginalized over.  However, as we
mentioned earlier, the fact that the likelihood function is
non-Gaussian in $r$ if $r$ is close to $\pm 1$ implies that the Fisher
matrix estimates may not apply.  Therefore, the following constraints
with marginalization over $r$ are only to be taken as rough
indicators.

We use moderate values for $r$ ($0.4$ for SZ and $-0.5$ for optical)
to compare the cosmological constraints assuming (1) fixed $r$ values,
and (2) $r$ to be a free parameter in the Fisher matrix.
Table~\ref{tab:constraints} contains our results for three of the
survey assumptions.  As can be seen, while the error bars for DES are
only slightly affected by marginalization over $r$, those for SPT
increase by a factor of 2 to 3.  The reason is again related to
the mass binning; since our fiducial SPT survey does not include mass
binning, there is no information about the shape of the halo mass
function, which, if present, can improve the constraints on the
scatter in the observable--mass relation. In the absence of this shape
information, the constraints on the scatter is modest, which means
that marginalization of $\OmegaDE$ and $w$ over the acceptable region
of the parameter space will reach areas with very large scatter.
Since those areas are highly sensitive to the effects of halo assembly
bias, the marginalized errors will be significantly larger.  In the
last row of Table~\ref{tab:constraints} (SPT5), we assume five narrow
observable bins for SPT with bin size $\Delta {\rm log}_{0}\Mobs =
0.2$. In this case, the dark energy constraints are barely degraded
after marginalizing over $r$.  Thus, mass binning is a crucial
component of the data analysis for both DES and SPT to maximize their
potential as cosmological probes.  Note that, in all cases, $r$ itself
cannot be well-constrained like other nuisance parameters.  Since the
dependence on $r$ only affects the sample variance but not the
abundance, the information for constraining $r$ is insufficient.

\begin{deluxetable*}{ccccccccccc}
\tabletypesize{\scriptsize}
\tablecaption{Self-Calibration Constraints. \label{tab:constraints}}
\tablewidth{0pt}
\tablehead{
     &   
    \multicolumn{4}{c}{Self-Calibration with Fixed $r$} &  & \multicolumn{5}{c}{Self-Calibration Marginalized over $r$}    \\
    \cline{2-5}  \cline{7-11}\\
   \colhead{Survey} &
    \colhead{$\OmegaDE$} &    
    \colhead{$w$} &
    \colhead{$\ln\Mbias$} & 
    \colhead{$\sigmalnM^2$} &
      &
    \colhead{$\OmegaDE$} &    
    \colhead{w} &
    \colhead{$\ln\Mbias$} & 
    \colhead{$\sigmalnM^2$} &
    \colhead{$r$}
    }
\startdata
SDSS	&	0.066	&	0.240	&	0.411	&	0.086	&	&	0.074	&	0.251	&	0.460	&	0.108	&	0.294	\\
DES	&	0.006	&	0.045	&	0.051	&	0.022	&	&	0.006	&	0.047	&	0.053	&	0.025	&	0.125	\\
SPT	&	0.010	&	0.076	&	0.104	&	0.028	&	&	0.025	&	0.177	&	0.355	&	0.149	&	1.300	\\
SPT5	&	0.009	&	0.061	&	0.079	&	0.017	&	&	0.010	&	0.062	&	0.087	&	0.027	&	0.357	\\ \enddata
\tablecomments{
Cosmological constraints with fixed cross-correlation coefficient $r$,
and with marginalized $r$.  We assume a WMAP3 cosmology, and the
nuisance parameters are the same as those in Fig.\@~\ref{fig:surveys}.
After marginalization over $r$, the constraints from binned cluster
samples (SDSS, DES, and SPT5) are barely degraded, while the
constraints from thresholded samples (SPT) are degrade by a factor of
2 to 3.  This result demonstrates the importance of mass binning.  In
all cases, $r$ cannot be well-constrained like other nuisance
parameters since it only affects the sample variance but not the
abundance.  We emphasize, however, that the second part of this table
are to be interpreted as rough indicators, since the likelihood
function may not be Gaussian in $r$.
}
\end{deluxetable*}


\section{Summary}
\label{sec:conclusions}

Self-calibration analysis in galaxy cluster surveys relies on the
dependence of the halo bias on mass to simultaneously constrain
cosmology and the cluster observable--mass distribution.  Recent work
has shown that halo bias is sensitive not only to halo mass, but also
to secondary parameters related to the assembly history.  Here we
consider the effect of halo concentration on the bias as a specific
case of the secondary parameters (generally termed assembly bias), and
show how it might affect self-calibration analyses.  In particular, if
halo selection depends on halo concentration, the observed clustering
amplitude of the corresponding cluster sample will deviate from that
of a random selection of clusters with the same mass distribution.
This deviation in the observed clustering amplitude can result in
biased inferences of cosmological parameters, depending on (1) the
amount of scatter between halo mass and the observational mass proxy,
and (2) the correlation between the mass proxy and halo concentration.
For current surveys like SDSS, the statistical uncertainty is still
sufficiently large that the systematic error due to assembly bias is
negligible.  On the other hand, for an SPT-like survey, the expected
small amount of intrinsic scatter between the SZ decrement and halo
mass suggests that the impact of assembly bias on parameter estimation
is negligible; however, if the projection effect results in higher
scatter in high density regions, assembly bias may have significant
impact.  For a DES-like survey, where the mass proxy is likely to have
considerably larger scatter, we estimate that assembly bias can
displace the recovered dark energy parameters from their true values
by about $1\sigma$.  For an LSST-like survey, this systematic error
can exceed $2\sigma$ in $w$.  In the last two cases, halo assembly
bias may need to be explicitly included in the cosmological analysis
to avoid biasing of the recovered dark energy parameters.  We
emphasize, however, that our analysis has assumed the specific
dependence of halo bias on halo concentration found by
\citet{Wechsler06}.  If this dependence is shown to be smaller at high
masses, if the correlation relating the observable mass proxy and halo
concentration can be shown to be small, or if observables that are
more tightly correlated with mass can be found, the effect will be
mitigated.  We have shown that binning in mass is crucial for both
optical and SZ surveys, as marginalization over this correlation
coefficient can increase the expected errors of dark energy parameters
by a factor of a few if we only use thresholded counts.

\acknowledgments 
We thank Marcos Lima, Michael Busha, David Rapetti, Doug Rudd, Gil
Holder, David Weinberg, and Andrew Zentner for useful discussions.  We
are grateful to the anonymous referee for helpful comments.  The IDL
contour plots are modified from the routine provided at
http://www.davidpace.com.  We also thank OSU CCAPP and KIPAC for hospitality
and support during our visits.  HW and RHW were supported in part by
the U.S. Department of Energy under contract number DE-AC02-76SF00515.
ER was supported by the Center for Cosmology and Astro-Particle
Physics (CCAPP) at the Ohio State University.


\appendix
\section{Biased Parameter Estimation from Incorrect Models}
\label{app:fisher-implementation}

In this section, we explicitly implement the modified Fisher matrix
formalism developed in \S\ref{sec:fisher} for the case in which both
$P_A(\vec x|\theta)$ and $P_B(\vec x|\theta)$ are Gaussian.  Let
$\vec\mu(\theta)$ and $\C(\theta)$ be the mean and covariance matrix
defining $P_A(\vec x|\theta)$ in model $A$, which is related to the
likelihood function; let $\vec\mu^B(\theta)$ and $\C^B(\theta)$ be the
corresponding quantities in model $B$, which represent the observed
data.  Note that $\vec\mu(\theta)$ and $\C(\theta)$ contain the model
parameter $\theta$ that we are trying to fit, while $\vec\mu^B$ and
$\C^B$ contain the true parameter value $\theta_t$.  The
log-likelihood function of model $A$ reads (up to a constant)
\begin{equation}
2\mathcal{L} = -2\ln L(\vec x|\theta)=\ln\det \C+(\vec x - \vec\mu)^T\C^{-1} (\vec x - \vec\mu)  \ .
\end{equation}
Taking the derivative with respect to $\theta$ and averaging over
$\vec x$, the maximum likelihood estimator $\hat\theta$ can be found
by solving
\begin{equation}
\avg{2\mathcal L_{,i}} = {\rm Tr}[\C^{-1}\C_{,i}(1-\C^{-1}\avg{\D})+\C^{-1}\avg{\D_{,i}}]|_{\theta = \hat\theta} =0 \ ,
\end{equation}
where $\avg{\D} = \C^B + (\vec\mu^B-\vec\mu)(\vec\mu^B-\vec\mu)^T$ and
$\avg{\D_{,i}} = -2\vec\mu_{,i}(\vec\mu^B-\vec\mu)^T$.  We then set
$\hat\theta = \theta_t + \delta\theta$, linearize this equation with
respect to $\delta\theta$, and solve for $\delta\theta$.

To proceed further, we focus on two simple examples of interest.  The
first example is the effect of assembly bias; model $A$ corresponds
the standard self-calibration, while model $B$ corresponds to
self-calibration with assembly bias.  In this case, model $B$ changes
the sample variance but not the mean; thus
$\vec\mu(\theta)=\vec\mu^B(\theta)$ for all $\theta$ values, but
$\C(\theta)\neq \C^B(\theta)$.  After linearizing with respect to
$\delta\theta$, the linear equations for $\delta\theta$ read
\begin{equation}
{\rm Tr}\{ \C^{-1}\C_{,i}(1-\C^{-1}\C^B + \sum_j\C^{-1}\C_{,j}\C^{-1}\C^B\delta\theta_j 
)\}+ 2 \sum_j\vec\mu_{,j}^T\C^{-1}\vec\mu_{,i}\delta\theta_j  = 0  \ .
\end{equation}
After solving the linear equations,  we obtain the parameter deviation $\delta\theta$  
\begin{equation}
\delta\theta_j =\sum_i (\bold{F}^{-1})_{ij} {\rm Tr}\{\frac{1}{2}{\C^{-1}\C_{,i}\C^{-1}(\C^{\rm B}-\C)}\} \ ,
\end{equation}
where
\begin{equation}
{F}_{ij} = \vec\mu_{,i}^T\C^{-1}\vec\mu_{,j} + \frac{1}{2}{\rm Tr}\{\C^{-1}\C_{,i}\C^{-1}\C_{,j}\}
\end{equation}
is the Fisher matrix of the Gaussian likelihood function.  Note that
the bias in the recovered parameters is proportional to the difference
between models $A$ and $B$.

Note that in analyzing the data generated by model $B$ using model $A$
changes not only the recovered parameters but also their error bars.
By performing a similar calculation, the modified Fisher matrix with
systematics now reads
\begin{equation}
{\tilde F}_{ij}=\vec\mu^{T}_{,i} \bold C^{-1} \vec\mu_{,j} + 
\frac{1}{2} {\rm Tr}[\bold C^{-1} \bold C_{,i}\bold C^{-1} \bold C_{,j}\C^{-1}\C^{\rm B}] \ .
\end{equation}
The error bar for all parameters estimated in model $A$ using the data
generated by model $B$ can be recovered by inverting ${\bf \tilde F}$.
However, in the case of counts-in-cells, the likelihood function is
not perfectly Gaussian; it is convolution of Poisson and Gaussian
\citep[see e.g.][]{LimaHu04,HuCohn06}.  The modified Fisher matrix
thus reads
\begin{equation}
{\tilde F}_{ij}=\vec\mu^{T}_{,i} \bold C^{-1} \vec\mu_{,j} 
+ \frac{1}{2} {\rm Tr}[\bold C^{-1} \bold S_{,i}\bold C^{-1} \bold S_{,j}\C^{-1}\C^{\rm B}]  \ .
\end{equation}

As a second example, we consider the case in which model $B$ changes
the mean but not the variance of the data. One example is the effect
of modified gravity on the weak lensing shear cross power spectrum
\citep[e.g.][]{HutererLinder07}.  Here model $A$ is the General
Relativity prediction, while model $B$ is the modified gravitational
prediction.  In this case, $\vec\mu(\theta) \neq \vec\mu^B(\theta)$
while $\C(\theta) = \C^B(\theta)$.  The linear equation for
$\delta\theta_j$ reads
\begin{eqnarray}
{\rm Tr}\{
{\sum_j\C^{-1}\C_{,i}\C^{-1}\C_{,j}\delta\theta_j} 
 \}
-2 (\vec\mu^{\rm B}-\vec\mu)^T\C^{-1}\vec\mu_{,i} +2 \sum_j \vec\mu_{,j}^T\C^{-1}\vec\mu_{,i}\delta\theta_j
 = 0 \ ,
\end{eqnarray}
which is equivalent to
\begin{equation}
\delta\theta_j=\sum_i ({\bf F}^{-1})_{ij}    \{  (\vec\mu^{\rm B}-\vec\mu)^T\C^{-1}\vec\mu_{,i} \} \ .  
\end{equation}
Our formalism thus provides a different and generalizable route of
obtaining the systematic error.


\bibliography{ms}

\end{document}